\documentclass[a4paper,dvips,12pt]{article}
\usepackage{axodraw}
\usepackage{array,tabularx,epsfig,mathrsfs,graphicx,rotating}
\usepackage{ifthen}
\usepackage{fancybox}
\usepackage{amsfonts}

\newcommand{\beq}{\begin{equation}}
\newcommand{\eeq}{\end{equation}}

\begin{document}
\vspace*{-10mm}

\begin{flushright}
Preprint SINP MSU 2006-03/802 \\
\end{flushright} 

\vspace*{3mm} 
\begin{center}
{\large \bf Real and Image Fields of a Relativistic Bunch  }

\vspace{5mm} 
{ \bf B. B. Levchenko\footnote{ e-mail: levtchen@mail.desy.de }}\\
{\it Skobeltsyn Institute of Nuclear Physics, Moscow State University }\\

\end{center}

\abstract{\small 
\vspace*{-7mm}
\noindent\begin{quote}

We derive  analytical expressions for  external  fields of a charged relativistic bunch 
with a circular cross section. 
At distances far from the bunch, the field reduces to the relativistic modified Coulomb 
form and in the near region, reproduce the external fields of a continuous beam. 
If the bunch is   surrounded by conducting surfaces, the bunch self-fields are modified. 
Image fields generated by a bunch between two parallel conducting planes are studied in 
detail. Exact summation of image fields  by the direct method invented by Laslett 
allows  the infinite series to be represented in terms of elementary trigonometric functions.

\end{quote}
}

\vspace*{5mm}
\section{\large Introduction}

In an accelerator, the charged beam is influenced by an environment (beam
pipe, accelerator gaps, magnets, collimators, etc.), and a high-intensity
bunch induces surface charges or currents into this environment. This
modifies the electric and magnetic fields around the bunch. There is a
relatively simple method to account for the effect of the environment
by introducing image charges and currents.

Over forty years ago Laslett \cite{Ltt}  analyzed the influence of the transverse 
space-charge phenomena,  due to image forces, on the instability of the coherent transverse 
motion of an intense beam.  
Methods of image fields summation are described in his paper \cite{Ltt}
which presented  some field  coefficients calculated for infinite parallel plate vacuum chambers, 
magnetic poles and vacuum chambers with elliptical cross sections and variable aspect 
rations. 
The resulting  image field was calculated only in the linear approximation and 
depends linearly on the deviations $\bar{x}$ and $x$  of the bunch center and the position of a test 
particle, respectively, from the axis. They act therefore like a quadrupole causing a coherent 
tune shift.
The approximation  used  is incorrect if the field observation point  $x$ is located far from 
the bunch or if  the bunch center is close to a conducting wall.

In the present paper we consider the problem  of the image field  summation once again
for  a very simple geometry, namely,   a relativistic bunch moving between infinitely wide  parallel 
conducting planes.
 The problem is far from being a pure academic one \cite{bbl}.  In applications, in particular by 
 study of the electron cloud effect \cite{ece} and the dynamics of photoelectrons in the beam
 transport system, 
 it is important to know the distribution of electromagnetic fields not only in vicinity of
  the bunch, but in  the whole gap.
 We did not find publications with attempts to sum up the series (\ref{a3-1}) in an
 approximation beyond the linear one. In  Section 3 we present the exact solution of 
 the problem.
 
 To solve the problem formulated above, in Section 2 we first derive an equation  for the external
 electromagnetic field generated by a cylindrical bunch of charged particles. The task is  specified 
 as follows.

The external radial electric $\vec{E}_{\perp}$, and azimuthal magnet $\vec{B}_{\phi}$  self fields 
for a round unbunched relativistic beam of the radius $b$ and a uniform  charge density are 
\cite{Wied}-\cite{Chao}
\begin{eqnarray}
 E_{\perp}&=&\kappa\frac{2q\lambda}{r},  \label{I-1}\\
 B_{\phi}&=&\frac{\mu_0}{4\pi}\frac{2q\lambda}{r}c\beta,
 \label{I-2}
\end{eqnarray}
where $\kappa= 1/4\pi\epsilon_0$,  $\lambda$ is the linear charge density,  $q$ is the 
charge,   $\beta = v/c$ is a normalized velocity of the beam constituents and $c$ is the velocity of light. 
In many applications,  equations  (\ref{I-1}) and (\ref{I-2}) are used to describe   fields of 
an individual bunch too. 
However, in the form (\ref{I-1}), (\ref{I-2}) the bunch fields  do not depend on the bunch 
energy and at large distances do not follow the Coulomb asymptotic.
This contrasts sharply with the fields produced (at $t=0$) by a rapidly moving  single 
charge $q$
\beq
 \vec{E}\,=\, \kappa\frac{q\,\gamma}{ r^2} \Big [ 
 \frac{1-\beta ^2}{1-\beta ^2\, sin^2\,\theta}\Big ]^{3/2}\,
\frac{\bf \vec{r}}{r},
\hspace*{5mm}
\vec{B}\,\sim \vec \beta \times \vec E,
\label{eq:2}
\eeq
where $\theta$ is the angle which the vector ${\bf \vec{r}}$ makes with the z-axis. 
Along the direction of motion the electric field is become weaker  
in $\gamma ^2$ times, while in the transverse direction the electric field is 
enhanced by the factor $\gamma$
\beq
 E_{\perp}\,=\,  \kappa \frac{q\,\gamma }{r^2} .
\label{I-4}
\eeq
Here, $\gamma$ denotes the particle Lorentz factor.

In the next section we derive an expression for the transverse component of the bunch electric field,
which  the defects indicated above are  rectified,
and find the conditions at which the bunch fields
are represented by (\ref{I-1}) and (\ref{I-2}).

\section{\large Self-Fields of a Charged Finite Cylinder \\
\ \ \ \ \ with a Circular Cross Section}

Let us consider a bunch of charged particles uniformly distributed with a  density $\rho$
within a cylinder of  length L and  an elliptical  cross section.  The ellipsoid
semi-axis in the x-y plane are $a$ and $b$ and the coordinate  z-axis is along the bunch axis.
Suppose that the bunch is moving along the z-axis with a relativistic velocity 
$\vec{v}=c\vec{\beta}$.

To compute the radial electric  field of such a rapidly moving  bunch, we have to
sum up fields of the type  (\ref{eq:2}), generated by the bunch  constituents. 
In this way we get \cite{FFM}
\beq
E_{\perp}(r,\xi,z)\,=\,\kappa\rho \gamma\big \{zI_1\,+\,(L-z)I_2\big \}
\label{a2-1}
\eeq
with
\begin{eqnarray}
I_1\,&=&\,\int\int\frac{(r-\sigma \cos(\xi-\phi))\,\sigma d\sigma d\phi}
{\big ( r^2+\sigma^2-2r\sigma \cos(\xi-\phi)\big )
\sqrt{\gamma^2 z^2+r^2+\sigma^2-2r\sigma \cos(\xi-\phi)}} 
\label{a2-i1}\\
I_2\,&=&\,\int\int\frac{(r-\sigma \cos(\xi-\phi))\,\sigma d\sigma d\phi }
{\big ( r^2+\sigma^2-2r\sigma \cos(\xi-\phi)\big )
\sqrt{\gamma^2 (L-z)^2+r^2+\sigma^2-2r\sigma \cos(\xi-\phi)}}
\label{a2-2}
\end{eqnarray}
where $\sigma$ is the distance in the x-y plane from the z-axis  to the elementary
charged volume and 
\beq
0<\sigma <\frac{ab}{\sqrt{a^2\sin^2 \phi + b^2\cos^2 \phi}},\ \ \ \ 0<\phi <2\pi.
\eeq
Equation (\ref{a2-1}) represents the radial electric field as observed at a distance $r$ 
from the bunch axis, at an angle $\xi$ relative to the $x$ axis and at a distance $z$
from the bunch tail.

The integrals $I_1$ and $I_2$ can  be estimated only numerically \cite{FFM}, if integrands
are taken as it is.  However,  the integrands  are easy to simplify if the bunch is relativistic, 
$\gamma \gg 1$, and   we would like calculate the field
in vicinity   of the bunch,  $r \sim L$, but at distances much larger than
the bunch radius, $b \ll r$.

  To simplify,  we make use of the notation
$$A=\frac{\sigma}{r},\ \ B=A \cos(\xi-\phi),\ \ Y=A^2-2B,\ \
C_1=\Big[1+\frac{\gamma^2z^2}{r^2}\Big]^{-1},\ \ X=C_1\cdot Y$$
and the integrand of $I_1$ can be written as
\beq
(r^2+\gamma^2z^2)^{-1/2}A(1-B)(1+Y)^{-1}(1+X)^{-1/2}\,.
\label{a2-3}
\eeq
Now we expand the above expression in a power series by using $A$ as a small parameter
and keeping only terms  up to the power $A^4$ at each step.  For the bunch shaped as
a circular cylinder, $a=b$ and we may set $\xi=0$. Due to the fact that
\beq
\int_0^{2\pi} \cos^{2k+1}\phi\,d\phi\,=\,0,
\eeq
all odd power of $B$ vanish after integration in $\phi$. This greatly simplifies the series 
generated from (\ref{a2-3}). After lengthy algebraic manipulations  with (\ref{a2-3}), we get
\beq
(r^2+\gamma^2z^2)^{-1/2}A\Big [1-(1+\frac{1}{2}C_1)A^2+(2+C_1+\frac{3}{2}C_1^2)B^2\Big ].
\label{a2-4}
\eeq
Substituting this expression in (\ref{a2-i1}), we get
\beq
I_1\,=\,\frac{\pi b^2}{r\sqrt{r^2+\gamma^2z^2}}\Big(1+\frac{3}{8}C_1^2\frac{b^2}{r^2}\Big ).
\label{s1-5}
\eeq
By changing $z^2$ to $(L-z)^2$ in (\ref{s1-5}), we obtain  for $I_2$ the following result
\beq
I_2\,=\,\frac{\pi b^2}{r\sqrt{r^2+\gamma^2(L-z)^2}}\Big(1+\frac{3}{8}C_2^2\frac{b^2}{r^2}\Big),
\label{s1-6}
\eeq
where $C_2=\Big[1+\gamma^2(L-z)^2/r^2\Big]^{-1}$. Notice that for particles uniformly 
distributed   in the bunch volume, $\rho=qN/\pi b^2L$, where N is number particles per bunch.
Substituting equations  (\ref{s1-5})-(\ref{s1-6}) in (\ref{a2-1}), finally we arrive to
\beq
E_{\perp}(r,z)\,=\,\kappa\frac{qN\gamma}{Lr}\Big\{\frac{z}{\sqrt{r^2+\gamma^2z^2}}\Big(1+ 
\frac{3}{8}\frac{b^2}{r^2}C^2_1\Big)+ 
\frac{L-z}{\sqrt{r^2+\gamma^2(L-z)^2}}\Big(1+ \frac{3}{8}\frac{b^2}{r^2}C^2_2\Big) \Big\}.
\label{s1-14}
\eeq 
This equation describe the electric field produced by a rapidly moving circular 
bunch.

The field of a  relativistic bunch described by (\ref{s1-14}),  has different behavior 
at distances far apart of the bunch and in the near region, $r \le L$. At very large distances,
$r \gg \gamma z$ and $r \gg \gamma (L-z)$,  equation (\ref{s1-14}) reduces to the Coulomb form
(\ref{I-4}). At the same time, in the near region and beyond the bunch tails, 
$\gamma z \approx \gamma (L-z)\gg r$  
and equation (\ref{s1-14}) simplifies to
\beq
E_{\perp}\,=\,\kappa\frac{2qN}{L}\frac{1}{r}, 
\label{s1-3}
\eeq 
which coincide with the external field  (\ref{I-1}) of a continuous beam with 
$\lambda = N/L$.

Similarly we can show that the azimuthal magnetic field of the bunch is 
\beq
B_{\phi}\,=\,\frac{\mu_0}{4\pi}\frac{\beta c}{\kappa}E_{\perp}(r,z).
\label{s1-16}
\eeq 

\section{\large Fields from Image Charges}


Following Laslett \cite{Ltt} (see also \cite{Wied}), we consider a relativistic bunch of the length
$L$ between infinitely wide   conducting planes at $x=\pm h$. Suppose that  constituents of the 
bunch are positively charged. For full generality, let 
the circular particle bunch be displaced in the horizontal plane by $\bar{x}$
from the midplane (0,y,z), and the observation point of the field be  at $(x,0,0)$
between the conducting parallel planes. The end points of the bunch are at $z=\pm L/2$.
The boundary condition for electric fields is $E_z(\pm h)=0$ on the conducting plane  
and is satisfied if the image charges change sign from image to image. 
Suppose that the distance between planes is of the order $L$.
Thus, the electric field
of each image is described by (\ref{s1-3}).
To calculate the image electric field  $E_{\perp, image} (x)$ in front of the plate, 
we add the contributions from all image  fields in the infinite series \cite{Ltt}
\begin{eqnarray}
\hspace{-4mm}
E_{\perp , image}(x,\bar{x})&\,=\,&\kappa\frac{2qN}{L}\cdot
\nonumber\\
&\Big\{& \hspace*{-3mm} (2h-x_1)^{-1}\ -(2h+x_1)^{-1}\ -(4h-x_2)^{-1}\ +(4h+x_2)^{-1}
\nonumber\\
&+& \hspace*{-3mm} (6h-x_1)^{-1}\ -(6h+x_1)^{-1}\ -(8h-x_2)^{-1}\ +(8h+x_2)^{-1}
\nonumber\\
&+& \hspace*{-3mm} (10h-x_1)^{-1}-(10h+x_1)^{-1}-(12h-x_2)^{-1}+(12h+x_2)^{-1}+...
\label{a3-1}
\Big\},
\end{eqnarray}
where $x_1=x+\bar{x}$ and $x_2=x-\bar{x}$. These image fields  must be added to the direct 
field of the bunch (\ref{s1-3}) to meet the boundary  condition
that the electric field enters conducting surfaces perpendicularly.

In the original paper \cite{Ltt}, the series (\ref{a3-1}) was summed up only in the linear
approximation in $x$ and $\bar{x}$,
\beq
E_{\perp , image}(x,\bar{x})\,=\, \kappa\frac{4qN}{L}\frac{\epsilon_1}{h^2}(2\bar{x}+x).
\label{s2-17}
\eeq
The coefficient $\epsilon_1=\pi^2/48$ is known as the Laslett coefficient (or form factor)
for infinite parallel plate vacuum chambers and magnetic poles.
The approximation used in (\ref{s2-17}) is incorrect if the deviation of the bunch center
from the axis is large ($\bar{x}\sim h$) or if the field observation point  $x$ is located far off 
 the bunch.  
 Therefore, below we present the exact solution of the problem.

In Appendixes A and B we prove that the exact summation of the series (\ref{a3-1}) gives
\beq
E_{\perp , image}(x,\bar{x})\,=\,\kappa\frac{4qN}{Lh}\Lambda(\delta,\bar{\delta}),
\label{s2-18}
\eeq
where the image field structure function $\Lambda$  depends only on normalized variables
$\delta = x/h$,  $\bar{\delta} = \bar{x}/h$ in the form
\beq 
\Lambda (\delta,\bar{\delta})=\frac{1}{2} \Big [\frac{\pi}{2}\cdot \frac{\cos(\frac{\pi}{2}\bar{\delta})}
{\sin(\frac{\pi}{2}\delta)-\sin(\frac{\pi}{2}\bar{\delta})} -\frac{1}{\delta -\bar{\delta}}\Big ].
\label{s2-35}
\eeq
In Appendix A it is shown that in the linear approximation  equation (\ref{s2-18}) recovers the part
(\ref{s2-17}) derived by Laslett.

We shall now estimate  values of the function  $\Lambda$  in several particular points.
If the  observation point of the field is located 
at the  plane, $x=h$, then    $\delta=1$ 
and the structure function  depends only on   the bunch center position between planes,
 $\bar{\delta}$. 
Thus,  from  (\ref{s2-35}) we get
\beq
\Lambda(1,\bar{\delta})= \frac{1}{2}\Big \{\frac{\pi}{2} 
\frac{1+\sin(\frac{\pi}{2}\bar{\delta})}{\cos(\frac{\pi}{2}\bar{\delta})}
 - \frac{1}{1-\bar{\delta}}\Big \}.
\label{s2-20} 
\eeq 
Equation (\ref{s2-20}) is singular at $\bar{\delta}\rightarrow 1$  and  shows that 
the conducting plane attracts the bunch with increasing force 
 with the bunch displacement from the midplane. This phenomenon, involving the transverse movement 
 of the bunch as a whole,  arises from image forces and could lead to a transverse instability.

For a bunch in the midplane, $\bar{\delta}=0$, the summed image field at the surface equals
\beq
E_{\perp , image}(h,0)\,=\,\kappa\frac{4qN}{Lh}\Lambda(1,0)\,=\,\kappa\frac{2qN}{Lh}
(\frac{\pi}{2}-1).
\label{A2-60}
\eeq

The image field (\ref{s2-18}) must be added to the direct 
field of the bunch (\ref{s1-3}) to meet the boundary  condition. Thus
\beq
E_{\perp , tot}(x, \bar{x})\,=\, E_{\perp , bunch}+ E_{\perp , image} = \,\kappa\frac{2qN}{Lh} 
\Big (\frac{1}{\delta} + \, 2\Lambda(\delta,\bar{\delta})\Big ).
\label{3.2-55}
\eeq
 For a bunch in the midplane, $\bar{\delta}=0$,
we find from  (\ref{3.2-55}) the expression of the  transverse  component of electric field generated
by a relativistic bunch  moving between wide conducting parallel planes 
\beq
E_{\perp , tot}(x, 0)\,=\,\kappa\frac{2 qN}{Lh} \cdot\frac{\pi/2}{\sin(\frac{\pi}{2}\delta )}.
\eeq
That is, at the surface, $\delta =1$,  
the field is enhanced by a factor $\pi/2$    due to the presence of the conducting planes.

Notice that in the linear approximation (\ref{s2-17}) the field gradient, 
$\partial E_{\perp}/\partial x$, 
is independent of position $x$. Thus the tune shift experienced by each particle in the bunch
 is the same (a coherent tune shift). However, the exact result (\ref{s2-18}) demonstrates that 
the coherence is violated and equation (\ref{s2-18}) allows us to estimate the accuracy of the  
 linear approximation.

\section{\large Magnetic Images}

In the above, we have used electrostatic images. Magnetic images can be treated in much the
same way. Let the ferromagnetic boundaries be represented by a pair of infinitely wide
parallel surfaces at $x=\pm g$. The magnetic field lines must enter the magnetic pole faces
perpendicularly.
For magnetic image fields we distinguish  between $DC$ and $AC$ image fields.
The $DC$ field penetrates the metallic vacuum chamber and reaches the ferromagnetic poles.
In case of bunched beams the $AC$ fields are of rather high frequency and we assume that they
do not penetrate the thick metallic vacuum chamber.

The $DC$ Fourier component of a bunched beam current is equal to twice the average beam current
$qc\beta\lambda B$ \cite{Wied}, where $ B$ is the the Laslett bunching factor. Thus,
\beq
B_{y,image,DC}(x,\bar{x})\,=\,-\frac{\mu_0}{4\pi}\frac{2qN\beta c}{L}  B \cdot \frac{4}{g}
\Lambda(\eta,\bar{\eta}),
\label{S4-1}
\eeq
where $\eta=x/g$ and $\bar{\eta}=\bar{x}/g$ and  the function $\Lambda$ is of the form 
(\ref{s2-20}).

The contribution of magnetic $AC$ image field due to eddy currents in vacuum chamber walls is
similar to electric image fields
\beq
B_{y,image,AC}(x,\bar{x})\,=\,\frac{\mu_0}{4\pi}\frac{2qN\beta c}{L}(1-B)\cdot\frac{2}{h}
\Lambda(\delta,\bar{\delta}),
\label{S4-2}
\eeq
where the factor $(1-B)$ accounts for the subtraction of the $DC$ component.

The magnetic image fields must be added to the direct magnetic field (\ref{I-2}) from the bunch
to meet the boundary condition of normal components at ferromagnetic surfaces. That is, 
the summary magnetic field  between the conducting planes is
\begin{eqnarray}
B_{y,tot}(x,\bar{x})&=&B_{y} +B_{y,image,DC} + B_{y,image,AC} \nonumber \\
&=& \frac{\mu_0}{4\pi}\frac{2qN\beta c}{L} \Big \{\frac{1}{x} + (1-B)\frac{2}{h}
\Lambda(\delta,\bar{\delta}) - B\frac{4}{g}\Lambda(\eta,\bar{\eta})\Big \}.
\label{S4-3}
\end{eqnarray}

\section{\large Summary}

We have derived an approximate expressions for   electric  (\ref{s1-14}) and magnetic (\ref{s1-16}) 
self-fields produced by a relativistic circular bunch with uniform charge density. 
They show that at distances  far from  the bunch the electromagnetic field  coinsides with the field 
generated by a point-like charged paricle. 
At the same time, in the near region and beyond the bunch tails, 
the fields  coincide with the external self-fields of a continuous beam (\ref{I-1})-(\ref{I-2}).

We re-analyzed the problem of summing the image fields generated by  a  bunch  of charged
particles moving with a relativistic velocity  between infinitely wide  parallel conducting planes.  
The exact solution of the problem represented by the structure function of image fields 
 $\Lambda$  (\ref{s2-35}) depending only of the normalized variables.

\vspace*{4mm}

{\bf \large Acknowledgments} 
\vspace*{3mm}

The author is grateful to  P. Bussey and E. Lohrmann  for  reading a paper draft, comments and  
useful discussions. This study is partially supported  by the Russian Foundation for Basic Research
under Grant no. 05-02-39028.

\vspace*{8mm}
{\bf \Large Appendices}
\appendix 
\section{ \large  Image Fields in Vicinity of a Bunch }

Here we derive the main formula (\ref{s2-35}).

Let split the contribution of all image fields  (\ref{a3-1}) given in braces 
into two parts, 
\begin{eqnarray}
& & (2h-x_1)^{-1}\ -(2h+x_1)^{-1}\ -(4h-x_2)^{-1}\ +(4h+x_2)^{-1} \nonumber\\
&+&  (6h-x_1)^{-1}\ -(6h+x_1)^{-1}\ -(8h-x_2)^{-1}\ +(8h+x_2)^{-1}
\nonumber\\
&+&  (10h-x_1)^{-1}-(10h+x_1)^{-1}-(12h-x_2)^{-1}+(12h+x_2)^{-1}+...
\label{A1-1}\\
&=& \sum_{k}^{\infty}\Pi_k (x_1,h) - \sum_{m}^{\infty}\Pi_m (x_2,h),
\label{A1-2}
\end{eqnarray}
where $\Pi_k$ represents the contribution from the  negative charged images and
 $\Pi_m$ is  the contribution from the positive charged images.
Here and hereinafter,  indexes $k$ and $m$ are possess odd, $k$=1,3,5,..., and even, 
$m$=2,4,6,... 
values.

An expansion of denominators of $\Pi_k$ and $\Pi_m$ into a power series of  small 
parameters $\delta_1=x_1/h < 1$ and  $\delta_2=x_2/h < 1$ gives
\begin{eqnarray}
\Pi_k (x_1,h)&=&\frac{1}{2kh-x_1}-\frac{1}{2kh+x_1}=\frac{2x_1}{(2kh)^2-x_1^2}=
\frac{2}{h}\sum_{n=1}^{\infty}\frac{\delta_1^{2n-1}}{(2k)^{2n}},    \label{A1-3} \\
\Pi_m(x_2,h)&=&\frac{1}{2mh-x_2}-\frac{1}{2mh+x_2}
 =\frac{2x_2}{(2mh)^2-x_2^2}=
\frac{2}{h}\sum_{n=1}^{\infty}\frac{\delta_2^{2n-1}}{(2m)^{2n}}.
\label{A1-4}
\end{eqnarray}
Now it is  evident that the space structure of the image fields between planes is 
characterized by a specific function $\Lambda (\delta_1,\delta_2)$,  we  term it  
the structure function,
\beq
\sum_{k}^{\infty}\Pi_k - \sum_{m}^{\infty}\Pi_m=\frac{2}{h}\Lambda (\delta_1,\delta_2).
\label{A1-20}
\eeq
with
\beq
\Lambda (\delta_1,\delta_2)=\sum_{k}^{\infty}\Big [
\frac{\delta_1}{(2k)^2}+\frac{\delta_1^3}{(2k)^4}+...\Big ]
-\sum_{m}^{\infty}\Big [ 
\frac{\delta_2}{(2m)^2}+\frac{\delta_2^3}{(2m)^4}+... \Big ].
\label{A1-21}
\eeq
The structure function $\Lambda$  depends only on the  normalized variables.
 
To proceed further, let us define the following auxiliary quantities
\begin{eqnarray}
M_j^{(-)}&=&\sum_{k}^{\infty}\frac{1}{(2k)^{2j}}-\sum_{m}^{\infty}\frac{1}{(2m)^{2j}}=
\sum_{n=1}^{\infty}\frac{(-1)^{n+1}}{(2n)^{2j}}=
\frac{1}{2^{2j}}\cdot\frac{(2^{2j-1}-1)\pi^{2j}}{(2j)!}|B_{2j}|,  \label{a23} \\
M_j^{(+)}&=&\sum_{k}^{\infty}\frac{1}{(2k)^{2j}}+\sum_{m}^{\infty}\frac{1}{(2m)^{2j}}=
\sum_{n=1}^{\infty}\frac{1}{(2n)^{2j}}=
\frac{1}{2^{2j}}\cdot\frac{2^{2j-1}\cdot\pi^{2j}}{(2j)!}|B_{2j}|,
\label{a24}
\end{eqnarray}
where $B_{2j}$ are Bernoulli numbers, $B_2=1/6$, $B_4=-1/30$, $B_6=1/42$ etc.
By adding and subtracting the leftmost parts of  (\ref{a23}) and 
(\ref{a24}), we express $k$ and $m$ numerical series of (\ref{A1-21}) in terms 
of $M_j^{(-)}$ and $M_j^{(+)}$.    Therefore, we get from (\ref{A1-21})
\beq 
\Lambda (\delta_1,\delta_2)=\frac{1}{2}\sum_{n=1}^{\infty}\Big [ 
(M_n^{(-)}+M_n^{(+)})\delta_1^{2n-1} + (M_n^{(-)}-M_n^{(+)})\delta_2^{2n-1}\Big ]
\label{a26}
\eeq
or after substituting  of (\ref{a23})-(\ref{a24}) in (\ref{a26}), we  find the following form
of  the  structure function generated by the  charged bunch,
\beq 
\Lambda (\delta_1,\delta_2)=\frac{1}{2}\sum_{n=1}^{\infty}\Big [ 
(2^{2n}-1)\delta_1^{2n-1} - \delta_2^{2n-1}\Big ] 
\frac{\pi^{2n}}{2^{2n}(2n)!}|B_{2n}|\, .
\label{A1-31}
\eeq
Using only the linear terms  we recover the  part derived by Laslett \cite{Ltt}
 (see equation (\ref{s2-17}))
\beq 
\Lambda (\bar{x},x,h)= \frac{1}{h}\cdot\epsilon_1 (2\bar{x}+x).
\eeq 
An inspection of (\ref{A1-31}) shows that the contributions of negative  charged images
are enhanced by the factor $2^{2n}-1$, as compared with the contributions from  the positive 
charged images. Equation (\ref{A1-31}) also shows that for $x$ in the bunch center, $\delta_2 =0$ 
and the contributions from the  positive charged images are vanish.

At the  final step, it is  possible to rewrite  the infinite series (\ref{A1-31}) in terms of elementary
trigonometric  functions. To do this, recall the relations between the Bernoulli numbers
and the trigonometric  functions \cite{RG},\cite{CL} 
\beq 
z\, \tan(z) = \sum_{n=1}^{\infty}\frac{(2^{2n}-1)(2z)^{2n}}{(2n)!}|B_{2n}|,\hspace{10mm}
z\, \cot(z) =1-\sum_{n=1}^{\infty}\frac{(2z)^{2n}}{(2n)!}|B_{2n}|.
\label{A1-33}
\eeq 
After some algebraic manipulations and the use of (\ref{A1-33}), we get from  (\ref{A1-31})
a new exact and compact expression  of the structure function
\beq 
\Lambda (\delta_1,\delta_2)=\frac{1}{2} \Big [ \frac{\pi}{4}\tan ( \frac{\pi}{4}\delta_1)\, +\,
\frac{\pi}{4}\cot (\frac{\pi}{4}\delta_2)\,-\,\frac{1}{\delta_2}\Big ].
\label{A1-34}
\eeq
Now, if  we recall that $\delta_1 = (x+\bar{x})/h = \delta +\bar{\delta} $ and 
$\delta_2 = (x - \bar{x})/h = \delta -\bar{\delta} $, we obtain

\beq 
\Lambda (\delta,\bar{\delta})=\frac{1}{2} \Big [\frac{\pi}{2}\cdot \frac{\cos(\frac{\pi}{2}\bar{\delta})}
{\sin(\frac{\pi}{2}\delta)-\sin(\frac{\pi}{2}\bar{\delta})} -\frac{1}{\delta -\bar{\delta}}\Big ].
\label{A1-35}
\eeq

At a first glance, equation (\ref{A1-34}) or (\ref{A1-35}) is singular at $\delta_2=0$ or
$\delta=\bar{\delta}$, respectively. However, as we already discussed right after (\ref{A1-31}),  
it is not the case. Starting once again from (\ref{A1-31})  with $\delta_2=0$ and account  
(\ref{A1-33}), we get formally
\beq
\Lambda(\bar{\delta},\bar{\delta})=\frac{\pi}{8}\tan \big (\frac{\pi}{2}\bar{\delta}\big ).
\label{3.1-32}
\eeq

Equations   (\ref{A1-31}),   (\ref{A1-34})   and  (\ref{A1-35}) were derived assuming $\delta <1$ and
$\bar{\delta} <1$. Therefore, one cast doubts on validity of (\ref{A1-35}) at $\delta \sim 1$,
near the conducting plane.
For that reason in the next section we  re-expand  the series  (\ref{A1-1}) into a power
series of new small parameters.

\vspace*{5mm}

\noindent
{\bf \large B \hspace{4mm} Image Fields in Vicinity of a Conducting Plane}

\vspace*{3mm}

A similar derivation is used to obtain the field structure near a conducting surface.
For the case under consideration we have to choose  new small parameters for the  expansion. 
Each bracket in (\ref{A1-1}) we represent in the form $(1\pm \Delta)^{-1}$ and 
expand in series, recalling that at the plane $x\approx h$,
$$\Delta_1=\frac{h-x_1}{h} \ll 1,\ \ \ \ \ {\rm and}\ \ \ \ 
\Delta_2=\frac{h-x_2}{h} \ll 1.$$
     In this way,
\begin{eqnarray}
(2kh-x_1)^{-1}&=& [(2k-1)h]^{-1}\Big [1+\frac{\Delta_1}{2k-1}\Big ]^{-1}=
\frac{1}{h}\sum_{n=1}^{\infty}\frac{(-1)^{n-1}\Delta_1^{n-1}}{(2k-1)^{n}}, \\
(2kh+x_1)^{-1}&=&[(2k+1)h]^{-1}\Big [1-\frac{\Delta_1}{2k+1}\Big ]^{-1}=
\frac{1}{h}\sum_{n=1}^{\infty}\frac{\Delta_1^{n-1}}{(2k+1)^{n}},  \\
(2mh-x_2)^{-1}&=& [(2m-1)h]^{-1}\Big [1+\frac{\Delta_2}{2m-1}\Big ]^{-1}=
\frac{1}{h}\sum_{n=1}^{\infty}\frac{(-1)^{n-1}\Delta_2^{n-1}}{(2m-1)^{n}},  \\
(2mh+x_2)^{-1}&=& [(2m+1)h]^{-1}\Big [1-\frac{\Delta_2}{2m+1}\Big ]^{-1}=
\frac{1}{h}\sum_{n=1}^{\infty}\frac{\Delta_2^{n-1}}{(2m+1)^{n}}.
\end{eqnarray}
Let us introduce the following auxiliary notations
\begin{eqnarray}
L_j^{(\pm)}&=&\sum_{k}^{\infty} \Big [\frac{(-1)^{j-1}}{(2k-1)^j}-\frac{1}{(2k+1)^j}\Big ]
\pm\sum_{m}^{\infty}\Big [\frac{(-1)^{j-1}}{(2m-1)^j}-\frac{1}{(2m+1)^j}\Big ], \label{A2-31}\\
L_{1,j}^{(-)}&=&\sum_{k}^{\infty}\frac{1}{(2k-1)^j}-\sum_{m}^{\infty}\frac{1}{(2m-1)^j}=
\sum_{n=1}^{\infty}\frac{(-1)^{n+1}}{(2n-1)^j},  \label{A2-32}   \\
L_{2,j}^{(-)}&=&\sum_{k}^{\infty}\frac{1}{(2k+1)^j}-\sum_{m}^{\infty}\frac{1}{(2m+1)^j}=
\sum_{n=1}^{\infty}\frac{(-1)^{n+1}}{(2n+1)^j},   \label{A2-33}     \\
L_{1,j}^{(+)}&=&\sum_{k}^{\infty}\frac{1}{(2k-1)^j}+\sum_{m}^{\infty}\frac{1}{(2m-1)^j}=
\sum_{n=1}^{\infty}\frac{1}{(2n-1)^j},   \label{A2-34}      \\
L_{2,j}^{(+)}&=&\sum_{k}^{\infty}\frac{1}{(2k+1)^j}+\sum_{m}^{\infty}\frac{1}{(2m+1)^j}=
\sum_{n=1}^{\infty}\frac{1}{(2n+1)^j}.
\label{A2-35}
\end{eqnarray}
By simple manipulations with seriess (\ref{A2-32})-(\ref{A2-33})  and (\ref{A2-34})-(\ref{A2-35}), 
it is easy to prove that
\beq
L_{2,j}^{(-)}= 1 - L_{1,j}^{(-)}, \hspace{10mm} L_{2,j}^{(+)} = L_{1,j}^{(+)} - 1.
\label{A2-36}
\eeq
From (\ref{A2-32})-(\ref{A2-36}) now  easy to find
\begin{eqnarray}
L_n^{(+)}+L_n^{(-)}&=&\big [(-1)^{n-1}-1 \big ]L_{1,n}^{(+)}+\big [(-1)^{n-1}+1 \big ] 
L_{1,n}^{(-)}, \label{A2-38} \\
L_n^{(+)}-L_n^{(-)}&=&\big [(-1)^{n-1}-1 \big ]L_{1,n}^{(+)}-\big [(-1)^{n-1}+1 \big ] 
L_{1,n}^{(-)} + 2.
\label{A2-39}
\end{eqnarray}
Let introduce the image field structure function $\Lambda$ in the way similar to (\ref{A1-20})
and rewrite $\Lambda$  in terms of notations (\ref{A2-31})
\begin{eqnarray}
\Lambda(\Delta_1,\Delta_2)&=&  \nonumber\\
&=&\frac{1}{2}\sum_{n=1}^{\infty}\Big \{ \Delta_1^{n-1}\Big [
\sum_{k}^{\infty} \Big (\frac{(-1)^{n-1}}{(2k-1)^n}-\frac{1}{(2k+1)^n}\Big )\Big ]\nonumber\\
&  &\ \ \ \ \ \ -\, \Delta_2^{n-1}\Big [\sum_{m}^{\infty}\Big (\frac{(-1)^{n-1}}{(2m-1)^n}-\frac{1}{(2m+1)^n} 
\Big )\Big ] \Big \}\nonumber\\
&=&\frac{1}{2^2}\sum_{n=1}^{\infty}\Big \{\Delta_1^{n-1}(L_n^{(+)}+L_n^{(-)}) -
\Delta_2^{n-1}(L_n^{(+)}-L_n^{(-)}) \Big \}.
\label{A2-37}
\end{eqnarray}
To perform the summation in (\ref{A2-37}), we have to split the series (\ref{A2-37}) into even, $n=2i$, 
and odd, $n=2i-1$ parts and substitute   (\ref{A2-38}), (\ref{A2-39}) 
in equation (\ref{A2-37})
\beq
\Lambda(\Delta_1,\Delta_2)=\frac{1}{2}\sum_{n=0}^{\infty}\Big [
(\Delta_1^{2n}+\Delta_2^{2n})L^{(-)}_{1,2n+1}-(\Delta_1^{2n+1}-\Delta_2^{2n+1})L^{(+)}_{1,2(n+1)}
-\Delta_2^{2n}-\Delta_2^{2n+1}\Big ].
\label{A2-40}
\eeq
With the help of  (\ref{A2-32}), (\ref{A2-34}) and \cite{RG} we find
\beq
L^{(-)}_{1,2n+1}=\frac{\pi^{2n+1}}{2^{2(n+1)}(2n)!}|E_{2n}|, \hspace{10mm} 
L^{(+)}_{1,2(n+1)}=\frac{(2^{2(n+1)}-1)\pi^{2(n+1)}}{2\cdot [2(n+1)]!}|B_{2(n+1)}|,
\eeq
and
\beq
\sum_{n=0}^{\infty}(\Delta_2^{2n}+\Delta_2^{2n+1})=\frac{1}{1-\Delta_2},
\label{A2-42}
\eeq
where $B_{n}$ and $E_{n}$ are Bernoulli and Euler numbers, respectively. 
Thus,
\begin{eqnarray}
\Lambda(\Delta_1,\Delta_2)&=&\frac{1}{2}\Big \{2\cdot\frac{\pi}{4}-\frac{1}{1-\Delta_2}
+\frac{\pi^2}{8}(\Delta_2-\Delta_1) \nonumber\\
&+&\sum_{n=1}^{\infty}\Big [(\Delta_1^{2n}+\Delta_2^{2n})L^{(-)}_{1,2n+1}-
(\Delta_1^{2n+1}-\Delta_2^{2n+1})L^{(+)}_{1,2(n+1)}\Big ]\Big \}.
\label{A2-52}
\end{eqnarray}
By using the  results obtained  in the previous section and  the  decomposition
\beq
\sec (x) = \sum_{n=0}^{\infty}\frac{|E_{2n}|}{(2n)!}x^{2n},
\eeq
equation (\ref{A2-52}) can be finally expressed in terms of  trigonometric functions,
\beq
\Lambda(\Delta_1,\Delta_2)= 
\frac{1}{2}\Big \{\frac{\pi}{4}\sec (\frac{\pi}{2}\Delta_1)\,-\, \frac{\pi}{4}\tan (\frac{\pi}{2}\Delta_1)  
\,+\,\frac{\pi}{4}\sec (\frac{\pi}{2}\Delta_2)\, + \,\frac{\pi}{4}\tan (\frac{\pi}{2}\Delta_2)\,-\,
\frac{1}{1-\Delta_2}\Big \}.
\label{A2-54}
\eeq
The structure function  $\Lambda(\Delta_1,\Delta_2)$,  as written in  (\ref{A2-54}),
looks very different from (\ref{A1-34}). However, it is not difficult to check that
by use of the relations
$\Delta_1=1-\delta_1$ and $\Delta_2=1-\delta_2$,  equation (\ref{A2-54})   
transforms in (\ref{A1-34}) or  (\ref{A1-35}).  

In this way we  ensure that  equations (\ref{A1-34}), (\ref{A1-35})  and (\ref{A2-54}) are  
correct and  represent exact summation of image fields generated by a charged bunch
between infinitely wide  conducting planes.

{}
\end{document}